\documentclass[manuscript, review = false, screen]{acmart}

\def\BibTeX{{\rm B\kern-.05em{\sc i\kern-.025em b}\kern-.08emT\kern-.1667em\lower.7ex\hbox{E}\kern-.125emX}}

\copyrightyear{2018}
\acmYear{2018}
\setcopyright{acmlicensed}
\acmConference[Woodstock '18]{Woodstock '18: ACM Symposium on Neural Gaze Detection}{June 03--05, 2018}{Woodstock, NY}
\acmBooktitle{Woodstock '18: ACM Symposium on Neural Gaze Detection, June 03--05, 2018, Woodstock, NY}
\acmPrice{15.00}
\acmDOI{10.1145/1122445.1122456}
\acmISBN{978-1-4503-9999-9/18/06}

\begin{document}

%
\title{A novel hand-held interface supporting the self-management of Type 1 diabetes}

%
\author{Robert Spence}
\email{r.spence@imperial.ac.uk}
\affiliation{%
  \institution{Imperial College London}
  \streetaddress{South Kensington}
  \city{London}
  \state{United Kingdom}
  \postcode{SW7 2AZ}
}

\author{Chukwuma Uduku}
\affiliation{%
  \institution{Imperial College London}
  \streetaddress{South Kensington}
  \city{London}
  \state{United Kingdom}
  \postcode{SW7 2AZ}
}
\email{chukwuma.uduku04@imperial.ac.uk}

\author{Kezhi Li}
\authornote{K. Li is the corresponding author.}
\affiliation{%
  \institution{Imperial College London}
  \streetaddress{South Kensington}
  \city{London}
  \state{United Kingdom}
  \postcode{SW7 2AZ}
}
\email{kezhi.li@imperial.ac.uk}

\email{chukwuma.uduku04@imperial.ac.uk}

\author{Nick Oliver}
\affiliation{%
  \institution{Imperial College London}
  \streetaddress{South Kensington}
  \city{London}
  \state{United Kingdom}
  \postcode{SW7 2AZ}
}
\email{nick.oliver@imperial.ac.uk}

\author{Pantelis Georgiou}
\affiliation{%
  \institution{Imperial College London}
  \streetaddress{South Kensington}
  \city{London}
  \state{United Kingdom}
  \postcode{SW7 2AZ}
}
\email{pantelis@imperial.ac.uk}

%
\renewcommand{\shortauthors}{R. Spence, et al.}

%
\begin{abstract}



The paper describes the interaction design of a hand-held interface supporting the self-management of Type1 diabetes. It addresses well-established clinical and human-computer interaction requirements.

The design exploits three opportunities. One is associated with visible context, whether conspicuous or inconspicuous. A second arises from the design freedom made possible by the user's anticipated focus of attention during certain interactions.
A third opportunity to provide valuable functionality arises from wearable sensors and machine learning algorithms. The resulting interface permits ``What if?'' questions: it allows a user to dynamically and manually explore predicted short-term (e.g., 2 hours) relationships between an intended meal, blood glucose level and recommended insulin dosage, and thereby readily make informed food and exercise decisions.
Design activity has been informed throughout by focus groups comprising people with Type1 diabetes in addition to experts in diabetes, interaction design and machine learning. The design is being implemented prior to a clinical trial.

\end{abstract}

%
%



%

%

%
\maketitle

\section{Introduction}

The prevalence of chronic disease has significant economic and social consequences, so a hand-held application allowing a patient to self-manage their condition has much to offer. The design of one such application, the ARISES app \footnote[1]{ARISES = An Adaptive, Real-time, Intelligent System to Enhance Self-care of chronic disease.} is the subject of this paper.

Type 1 diabetes (T1DM) is a chronic condition characterized by insulin insufficiency occasioned by the autoimmune destruction of pancreatic beta cells \cite{oviedo2017review}. Subcutaneously administered insulin replacement therapy is the mainstay of treatment, and can be delivered as multiple daily injections or via a continuous subcutaneous insulin infusion pump. Diabetes health applications have been shown to successfully improve treatment outcomes (e.g., blood glucose control), health behaviour (e.g., self-monitoring of blood glucose), patient self-confidence, and patient satisfaction \cite{Bonoto-EffOfMob2017}. However, a study \cite{Connell-23ofUsers2017} from the leading mobile engagement platform found that a quarter of users abandon apps after just one use, with poor software user experience being a major barrier in mobile health application penetration. Therefore, a primary objective when developing the health app interface described in this paper was to ensure usability across a wide demographic without compromising the following;

1. Manual data input (e.g. of meal selection and exercise type and intensity);

2. Automated input of physiological parameters from wearable devices (e.g., blood glucose levels and heart rate);

3. Presentation, to the user, of recommended treatment (e.g., insulin dosage and risk aversion strategies);

4. Real-time visual presentation of predicted outcomes that would follow from treatment, recommended by machine learning algorithms;

5. Manual dynamic exploration of interrelations between relevant parameters (e.g., carbohydrate values, predicted blood glucose and insulin recommendation).

\section{Relevant Disciplines}
Two disciplines were particularly relevant to the design activity. A major one is human-computer interaction, the focus of this paper. Another is that of the metabolic sequelae associated with T1DM which impose specific clinical requirements upon the functionality of the interface. The latter clinical background, together with a summary of related requirements, is concisely addressed in Appendix A1, as are implications of the human visual system in Appendix A2 and user requirements in Appendix A3.  The detailed influence of those requirements is treated as and when appropriate within the discussion of human-computer interaction issues in the main body of the paper.

\section{DESIGN APPROACH}
The approach adopted for the development of the ARISES app followed the familiar route of a divergent brainstorming session followed by gradual convergence in which very extensive and regular interaction occurred between interaction designer and clinicians, machine learning specialists and potential users.

The principal and anticipated beneficial outcome of the design process followed the exploitation of three opportunities:

\textbf{Context}\\
The discipline of human-computer interaction already recognizes the importance and potential of visible context (e.g., \cite{pirolli-visual2001,findlay-active2003}).  Consideration of the cognitive and perceptual benefits arising from visible context, whether that context be conspicuous or inconspicuous, allowed potentially useful and novel aspects of the interface to be proposed.

\textbf{Attention}\\
During many brief interactions (e.g., the selection of a meal) a user need only pay attention to very few features of an interface, and should not be distracted by other features that are temporarily irrelevant: nevertheless, at the same time, the benefits of visible context should be maintained. That assumption can lead to the whole display area being available for items (e.g., menus) that would otherwise be restricted to a much smaller area and induce visual stress or distraction.

\textbf{Technology}\\
The use of wearable sensors (e.g., of blood glucose level, heart rate and temperature), combined with machine-learning technology, enables future blood glucose levels to be predicted with acceptable accuracy \cite{hayeri-predicting2018,Perez-DecisionSup2018, Li-ADLPlatform2018}. In turn, this provides an opportunity for a user to engage in smooth and continuous manual exploration of the relationship between an intended meal, predicted blood glucose and recommended insulin dosage.  The instantaneous display of predictions allows a considered but rapid decision to be made by a user.

\begin{figure}
\centering
\begin{minipage}{.49\textwidth}
  \centering
  \includegraphics[height=2.5in]{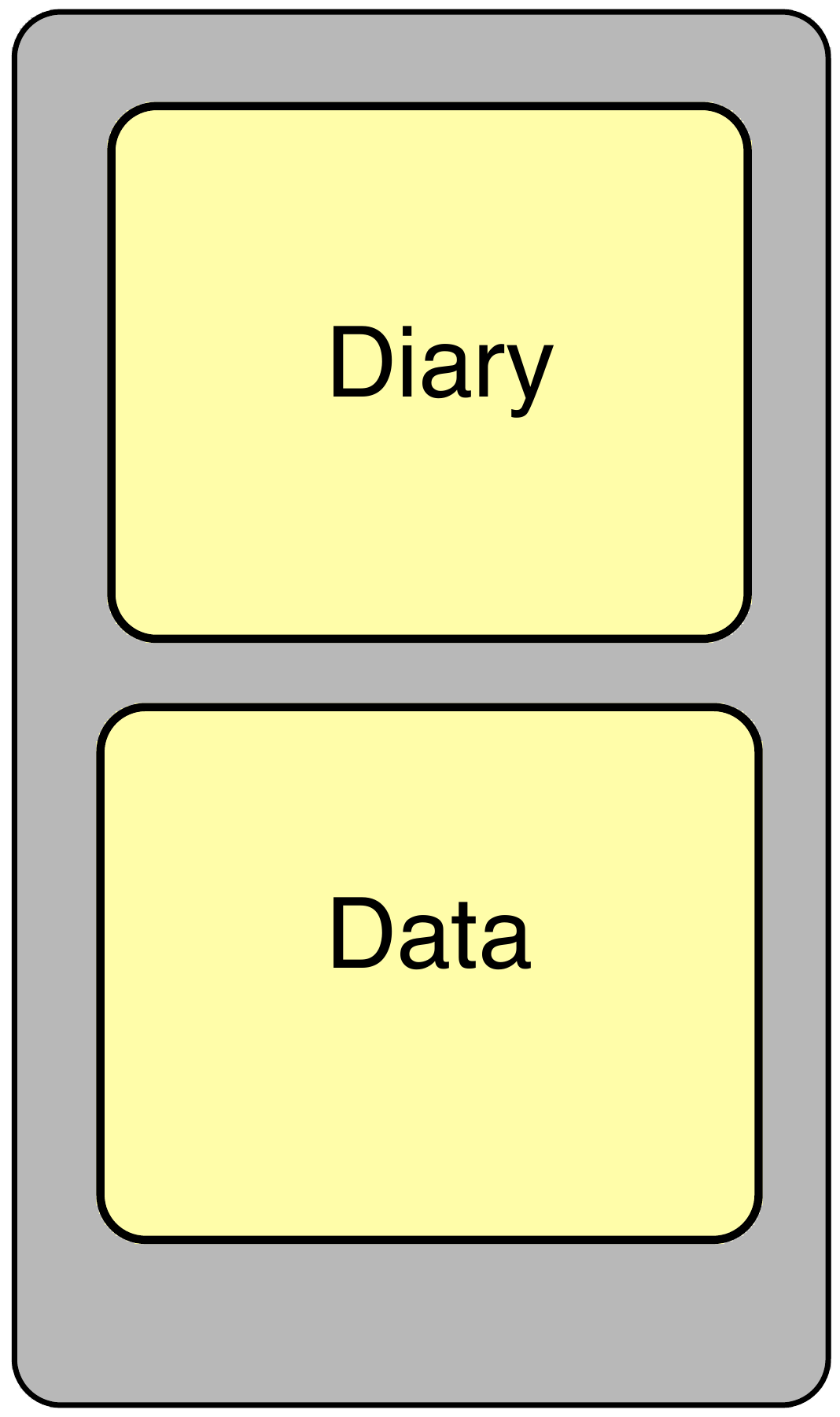}
  \captionof{figure}{The allocation of display area to temporal data (top)  and static data input and output (lower).}
  \label{fig:test1}
\end{minipage}%
\ \ \ \
\begin{minipage}{.49\textwidth}
  \centering
  \includegraphics[height=2.5in]{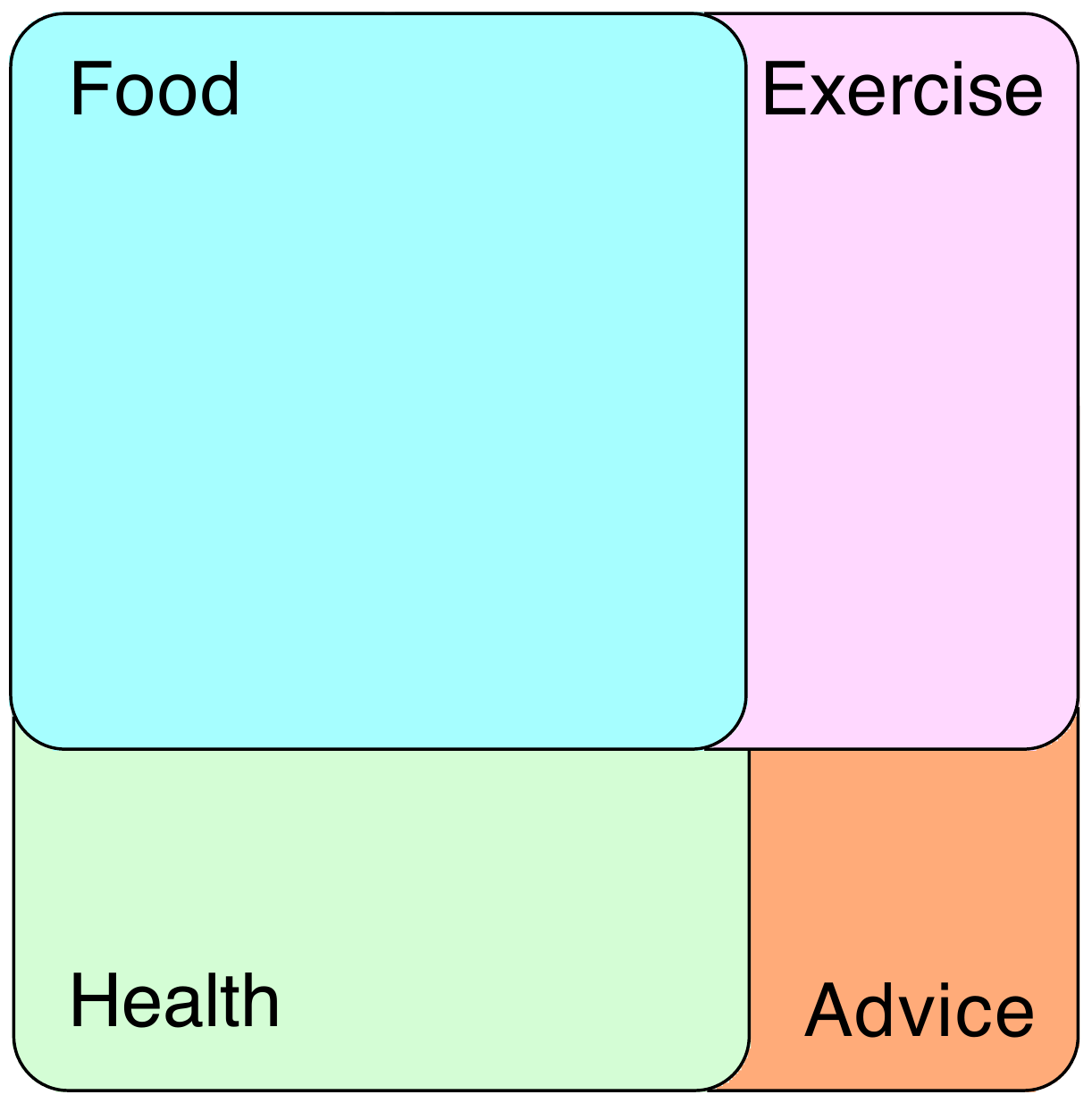}
  \captionof{figure}{Stacked layout of the four data regions occupying the lower part of the hand-held display. }
  \label{fig:test2}
\end{minipage}
\end{figure}



\section{Clinical Considerations}

Following a detailed review by specialist diabetologists of current diabetes decision support systems \cite{Li-ConvRecNN2018,breton-continuous2018,Herrero-AdvInsBolus2015}, five groupings of data required from, and to be presented to, the user were identified.  These were:

(1)	The graphical presentation of \textbf{blood glucose} data and access to continuous historical data on a displayed diary. Time-stamped meal and exercise data provide context for variations in blood glucose levels.

(2)	\textbf{Food}: parameters such as carbohydrate (`carb'), protein and fat values as well as alcohol intake.

(3)	\textbf{Exercise}: A categorical choice of planned exercise type and intensity.

(4)	\textbf{Health}: Ability for the machine learning system to identify and adapt its algorithm in response to stress and illness as well as to access recorded customizable parameters significant to the management of diabetes. These include filter-adjusted blood glucose measurements, and stress and illness, among others.

(5)	\textbf{Advice}: Presentation of trends associated with negative outcomes on blood glucose levels. The provision of both preventative advice and context via events displayed on a diary.

The interface design developed and described below takes account of these requirements.

\section{USER CO-DESIGN}

An important feature of the development of the interface described below was the involvement of potential future users (e.g. \cite{steen-human2012}). At regular semi-structured `focus group' meetings about 10 individuals with type-1 diabetes  met with clinicians, engineers and specialists in human-computer interaction and constituted a forum providing feedback regarding data input and presentation. A summary of such co-design is provided in Appendix A3: detailed references, where appropriate, are made in the main text.

\section{DESIGN DETAIL}

\subsection{Layout}

The management of type-1 diabetes requires the input, display, interpretation and exploration of two types of data: temporal and static. To maintain both types of data in useful visual context, the display real estate on the hand-held display was allocated as shown in Figure 1.  As discussed below, to exploit the advantages of context visibility, fixed parts of both the Diary and the static data area were designed to be always visible.

\subsection{Static Data}

The metaphor of stacked partially overlapping regions (Figure 2) is employed to allow the input and presentation of variables associated with the four principal data types. Any region can be brought to the `top' by a single touch, but its size is such that designated fixed parts of the other 3 regions are always available to provide visual context \footnote[2]{The permanently visible parts of the four regions are not primarily intended as `tags', although touch on one moves the corresponding region to `the top'. Their role is primarily to provide contextual information: for example to allow awareness of planned exercise while choosing a meal in the Food region.}. During such interaction the stacking order of the remaining regions is maintained.  It is expected that the nature of such interaction allows a user to develop a simple and intuitive mental model of interaction.

The sketch of Figure 3 illustrates the perceived value of context: while values of Food parameters are entered, important earlier/concurrent choices regarding Exercise are visible as a reminder.

\begin{figure}
\centering
\begin{minipage}{.5\textwidth}
  \centering
  \includegraphics[height=1.5in]{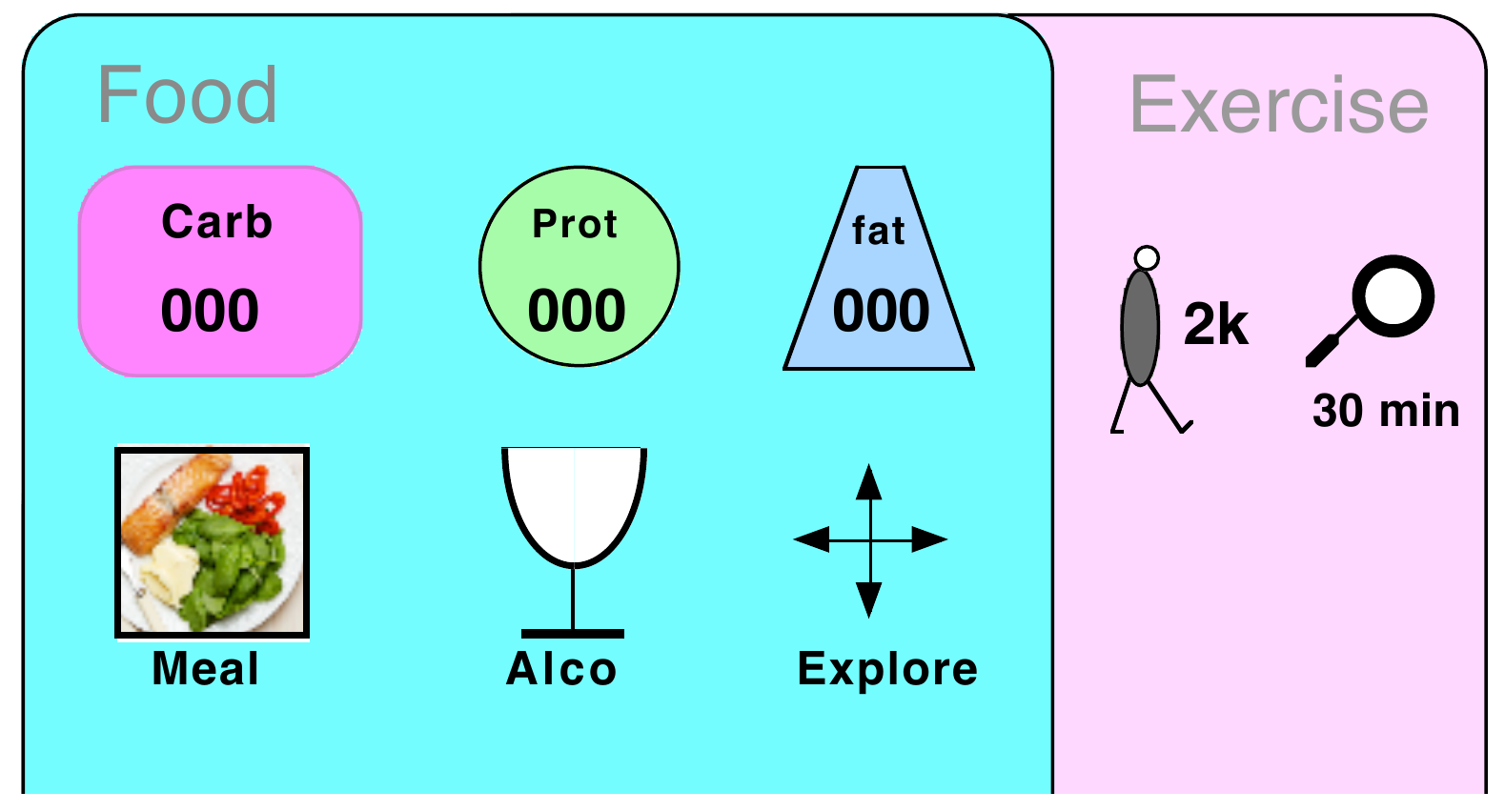}
  \captionof{figure}{Examples of the data concerning Food and Exercise that  \mbox{} \mbox{} the user can enter. Note that part of the Exercise region is always  \mbox{} \mbox{} visible while data concerning Food is being provided.}
  \label{fig:test3}
\end{minipage}%
\begin{minipage}{.5\textwidth}
  \centering
  \includegraphics[height=1.38in]{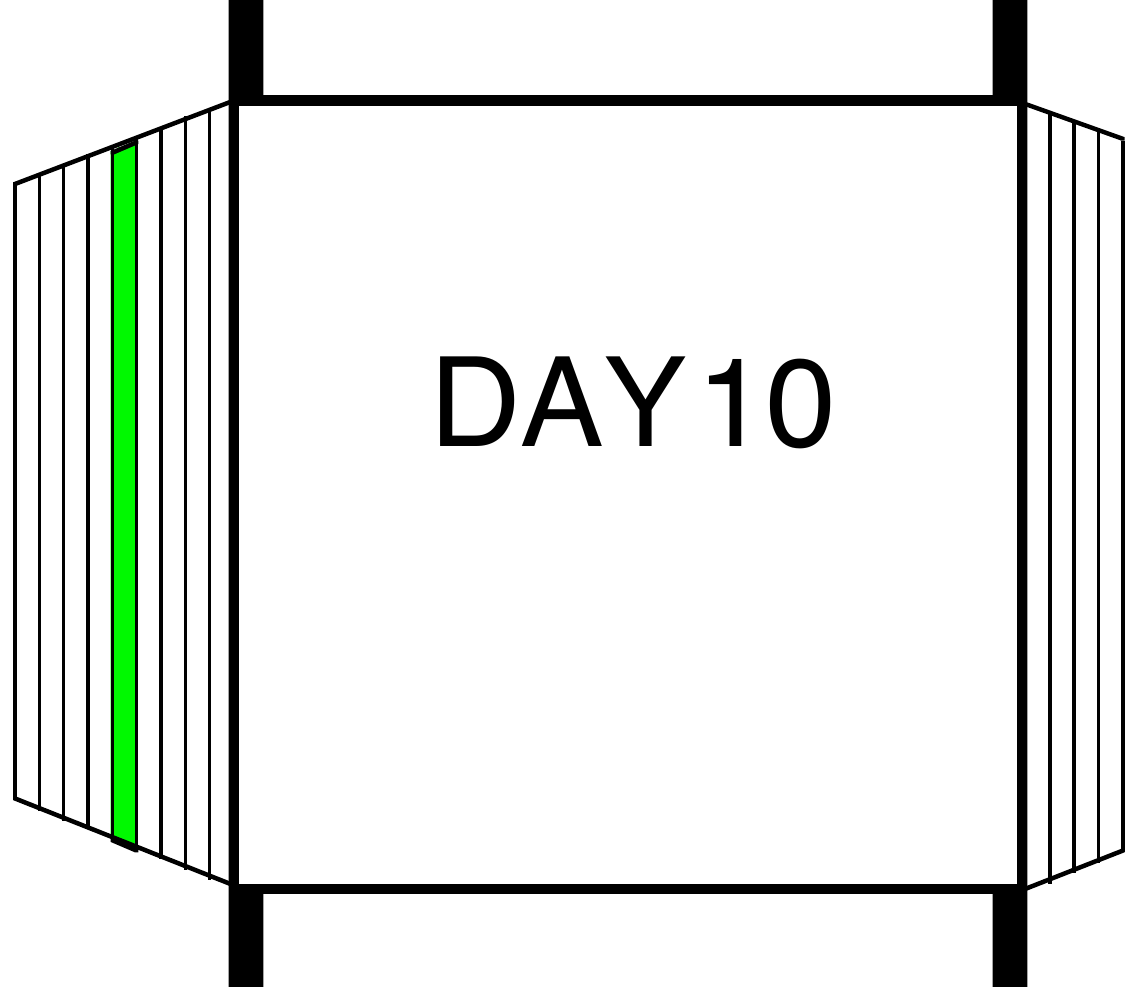}
  \captionof{figure}{Diagrammatic illustration of the `distortion' associated with the metaphor of the Bifocal Diary.}
  \label{fig:test4}
\end{minipage}
\end{figure}



\subsection{Temporal data}


Temporal data is one key to the successful management of Type-1 diabetes. It includes the choices (e.g., of food and exercise) made over past hours and days, as well as automatically captured physiological data such as blood glucose and insulin dose. To reflect the importance of temporal data the upper part of the interface (see Figure 1) is dedicated to a diary within which much of that data can be viewed and interacted with.

Given the limited display area available, and to cater for a user's interest in both quantitative data (e.g., todays' recorded and predicted blood glucose levels) and qualitative data (e.g., earlier occasions when a given meal was chosen), the metaphor underlying the Bifocal Display \cite{Spence-DataBaseNav1982,furnas-generalized1986,Mackinlay-ConeTrees1991,Card-Comm2012} was employed (Figure 4).  The `distortion' \cite{forlines-dtlens2005,leung-review1994} created by the metaphor, which has many applications (e.g., \cite{rao-table1994,bederson-datelens2004}) allows detailed observation of data variation (both recorded and predicted) within a single day (see the current day's blood glucose variation in the focus region of the Bifocal Diary in Figure 5) as well as useful qualitative encoding of events associated with adjacent days. Should one of the latter events be of interest, simple scrolling will lead to the appropriate day `snapping' into the `focus region' to allow detailed quantitative examination.

\begin{figure}
\includegraphics[height=2.7in]{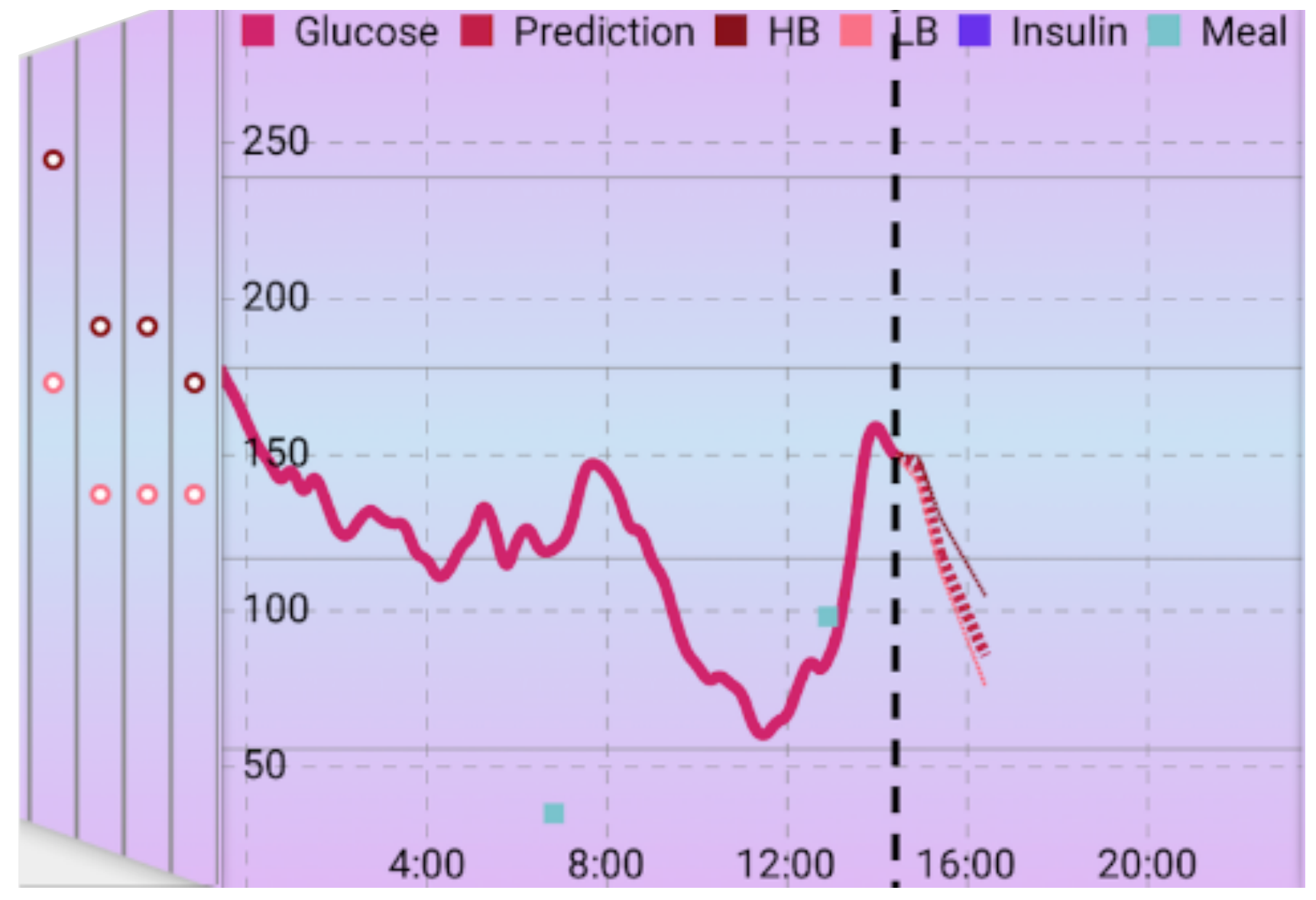}
\caption{The actual (before the vertical black line denoting `Now') and predicted variation of blood glucose, as well as its confidence limits.}
\end{figure}

\subsubsection{The Diary}
Figure 6 illustrates the qualitative encoding of data \cite{Mackinlay-ConeTrees1991,Card-Comm2012,rao-table1994} that can be presented in the limited (`distorted') regions of the diary. Bars indicate the percentage of a day that blood glucose levels were located within three regions (normal, hypoglycaemic and hyperglyceamic).  Simple but noticeable icons can show, for example, when a meal being planned was last consumed. There is potential for additional events to be added and for encoding freedoms such as vertical position to be explored \cite{Spence-DataBaseNav1982,Mackinlay-ConeTrees1991}.

Figure 7 illustrates the information that is presented in the focal region \cite{Spence-DataBaseNav1982,Card-Comm2012} for a single day.  In addition to recorded blood glucose level (red curve) and insulin doses applied (numbers in circles) other data appears in response to touch anywhere on the diary (and disappears on further touch). That additional data allows a user to recall decisions made concerning carb value together with meal images when chosen. Legibility will be enhanced if the hand-held device is rotated by 90 degrees.

The above decisions regarding temporal and static data led to the appearance of the `launch state' of the ARISES app as shown in Figure 8 \footnote{The term `launch state' is used in preference to `home state' with its frequent connotation of navigational difficulties. If a term such as `home state' is preferred the adjective `nomadic' might be appropriate.}.

\begin{figure}
\centering
\begin{minipage}{.39\textwidth}
  \centering
  \includegraphics[height=2.14in]{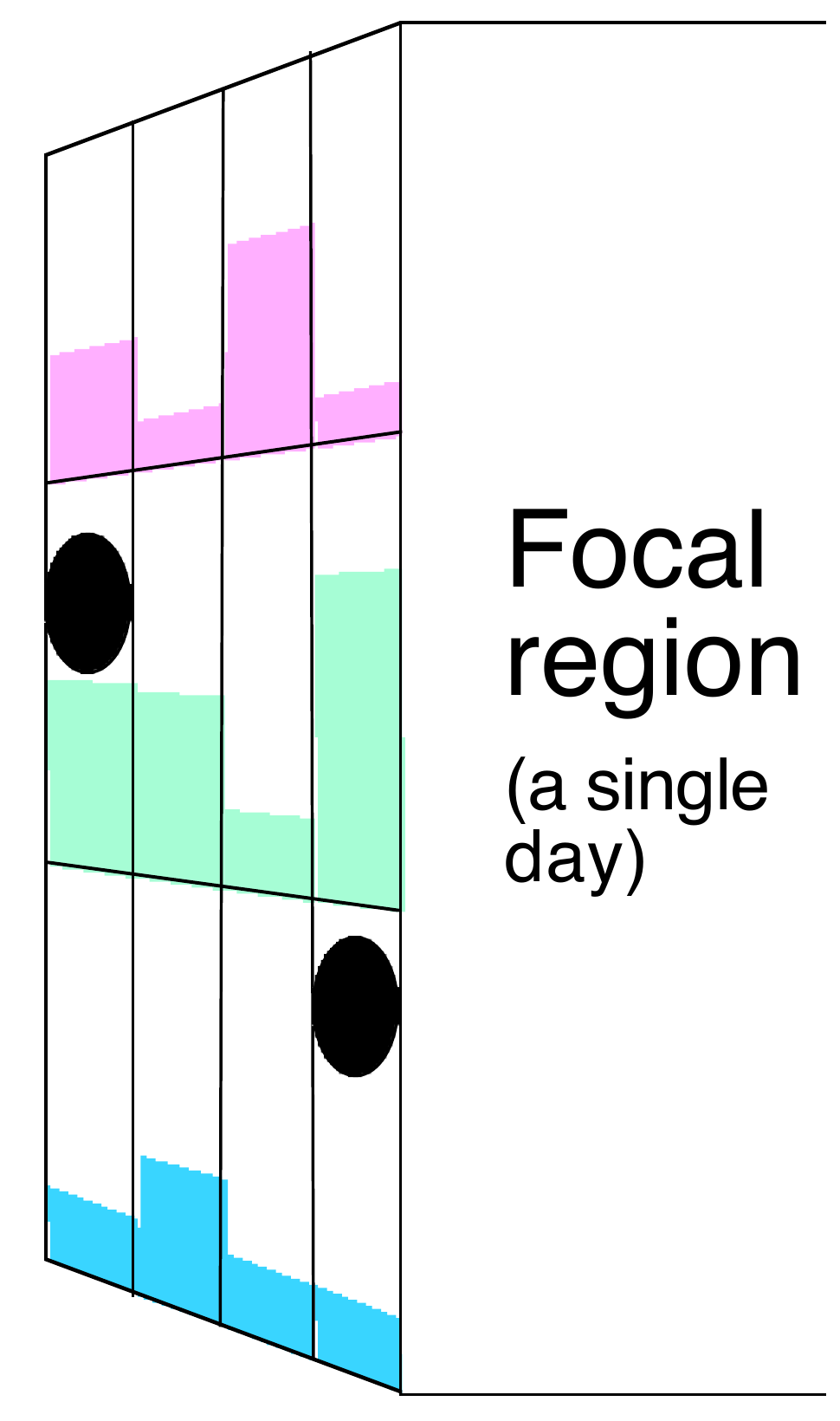}
  \captionof{figure}{For four days previous to that  which occupies the focal region, \mbox{} \mbox{} bars show the percentage of time that blood glucose level has stayed in three important regions. Black icons indicate important events, but vertical position can also encode a variable of interest.}
  \label{fig:test6}
\end{minipage}%
\ \ \
\begin{minipage}{.59\textwidth}
  \centering
  \includegraphics[height=2in]{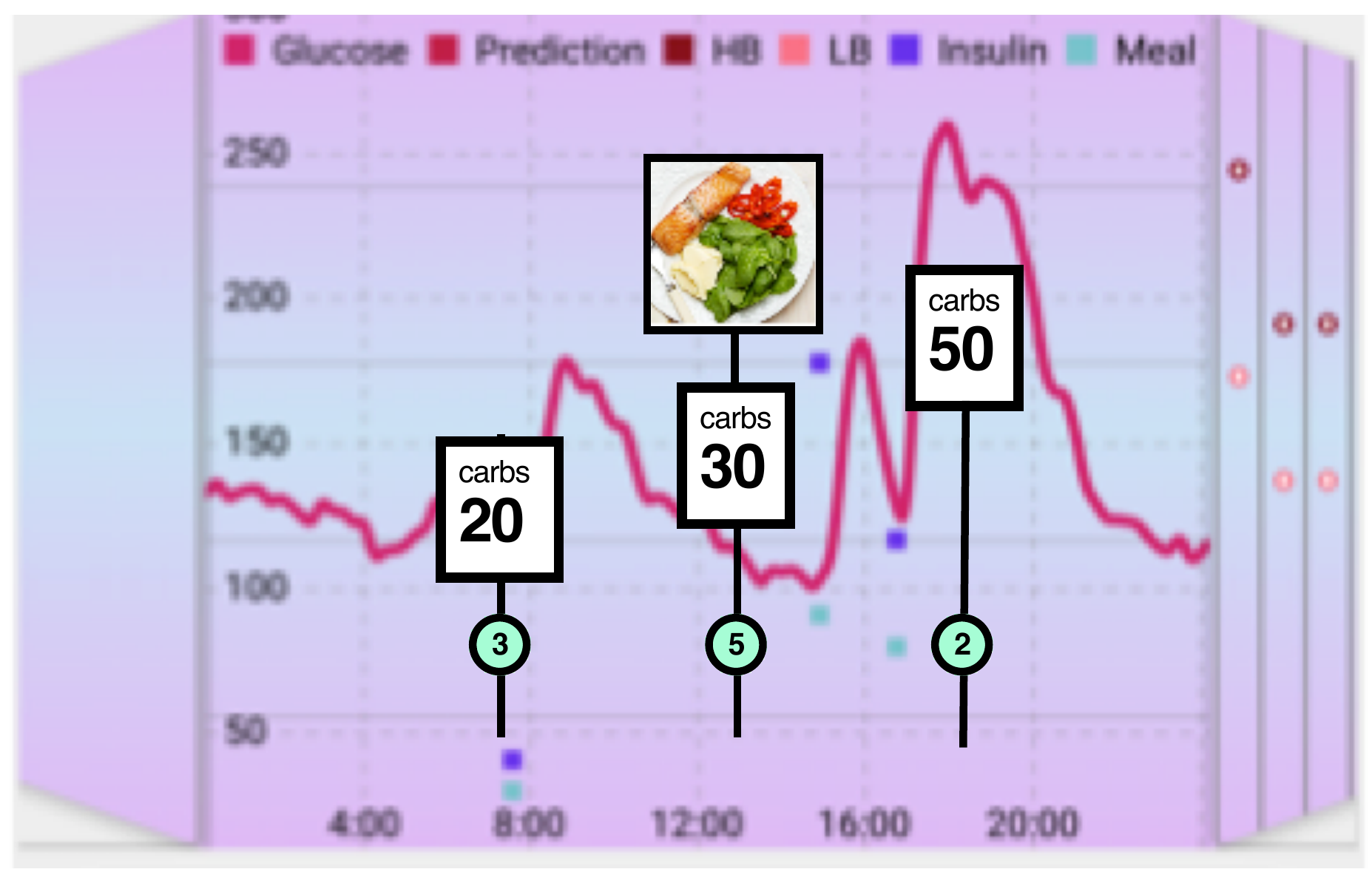}
  \centering
  \captionof{figure}{Example of the focal region content for a past full day. Recorded blood glucose is shown in red, and insulin doses delivered are shown in small circles with a blue background. The remainder of the presentation occurs when the diary is touched (anywhere), and shows carb values of meals consumed and, where available, a selected meal image.}
  \label{fig:test7}
\end{minipage}
\end{figure}

\subsection{Data input}
The manual input of data can be eased by recognizing one simple fact: that during the short period of time (typically less than 3 seconds) when data is being provided by a user (say, via a menu), that user's attention is primarily and often entirely directed to very few items on the display. Their attention is therefore less likely to be distracted if, for that very short time period, all other features of the display are rendered inconspicuous, leaving the entire area of the display free to accommodate menu items. Below we describe three illustrative ways in which this simple principle can be applied to a variety of data input activities.

\subsubsection{Numerical data input}

Figure 9 shows the interface designed to allow the selection of carbohydrate values associated with an intended meal. Note that items irrelevant to menu selection are not removed, but rather made inconspicuous in order to maintain context.  Also, the large size and spacing of menu items are implemented in response to end user feedback to support the population of visually impaired individuals with diabetic retinopathy \cite{Gella-Impairment2015,Tan-FactorsAss2017,Wu-MobileApp2017} and others with tremor.

\subsubsection{Categorical data input}

The same principle can benefit the input of categorical data. We take an example arising from the observation that people are creatures of habit, and often choose meals consumed on past occasions.  The interface shown in Figure 10 provides a scrollable collection of food images relevant to a default choice (here, Meal) corresponding to the time of day at which the meal image icon in the Food region is selected by touch, a choice that is easily changed. Following a meal selection, the image appears (Figure  11) surrounded by associated values of carbs, protein and fat and advised insulin requirement. Touch on the meal image selects that meal, incorporates its image in the Food region, and returns the app to its launch state (Figure 8).

\subsubsection{Exercise specification}

Habituation also characterizes a user's choice of exercise type and intensity, but the approach illustrated for meal choice in Figure 10 is inappropriate for exercise. Whereas meal selection involves a choice of one from many, any exercise type is chosen together with an intensity level that might vary from day to day. Also, whereas there might easily be a collection of 30 favourite meals, justifying the interface of Figure 10, the range of exercise types relevant to a given user might be very few. For this reason the interface shown in Figure 12 is proposed. One touch on the Exercise region displays a default menu based on history, and allows intensity to be adjusted following identification of exercise type by touch. Many scenarios involve very few touches, and the occasional but planned 2-week skiing holiday, for example, can easily be scrolled in (suggested by the greyed-out fraction of types at the top and bottom of the exercise list in Figure 12). The initial customization of default types will be achieved within a Settings mode that provides a comprehensive collection of exercise types and intensity levels.

\begin{figure}
\centering
\begin{minipage}{.49\textwidth}
  \centering
  \includegraphics[height=4.1in]{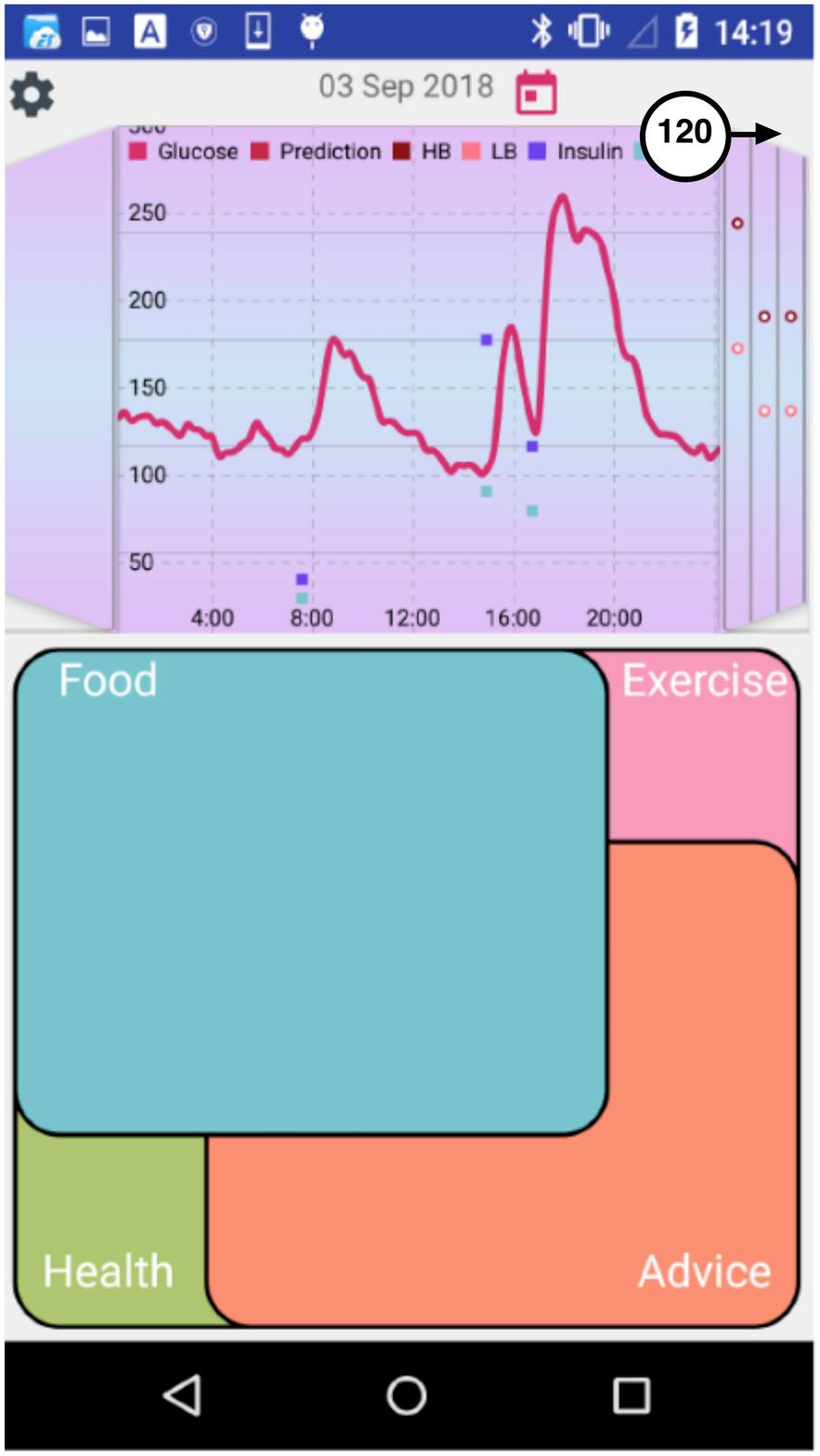}
  \captionof{figure}{The `launch' state of the ARISES app.}
  \label{fig:test8}
\end{minipage}%
\ \ \
\begin{minipage}{.49\textwidth}
  \centering
  \includegraphics[height=3.9in]{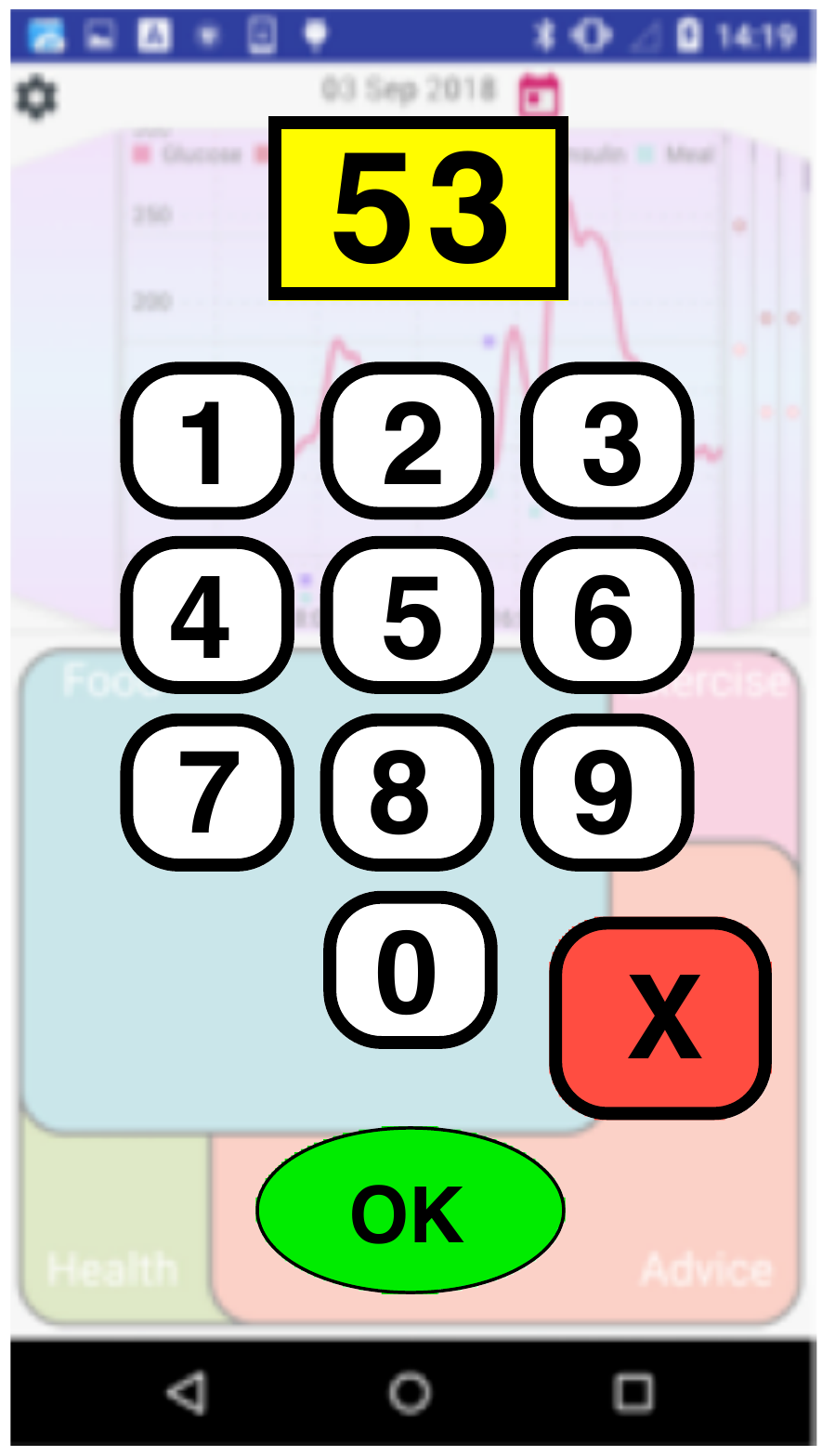}
  \centering
  \captionof{figure}{Menu options are more easily selected if made large. During carb value selection (taking, say, 500 to 2000 milliseconds), inability to see ghosted items would not be serious. }
  \label{fig:test9}
\end{minipage}
\end{figure}


\subsection{Dynamic Exploration}

User trepidation surrounding the self-management of T1DM stems largely from uncertainty about how blood glucose levels will change in response to daily choices such as meals, exercise and recommended insulin dosage. What the user would like to do is ask a ``what if?'' question such as ``If I increased the carb value of the meal I am just about to eat, how will that affect my predicted blood glucose over the next two hours and the recommended insulin dosage? ''

To enable such exploration, continuous wearable glucose monitoring technology records blood glucose levels every five minutes \cite{us2016approval}. The application of advanced machine learning algorithms then enables the ARISES app to make a prediction of blood glucose levels over the next two hours for a variety of hypothetical scenarios. That prediction can be essentially immediate, enabling the user to continuously vary a carb value and immediately see its effect, a process known as dynamic exploration  \cite{Spence-Graphical1971, Spence-InfoVis2014,spence-responsive1995, Neufeld-ExtendingTheDim2008}.  It would additionally be beneficial if the recommended insulin dosage can also be varied by manual sliding of the insulin recommendation and lead, for a given carb value, to a predicted blood glucose variation.

\begin{figure}
\centering
\begin{minipage}{.49\textwidth}
  \centering
  \includegraphics[height=4in]{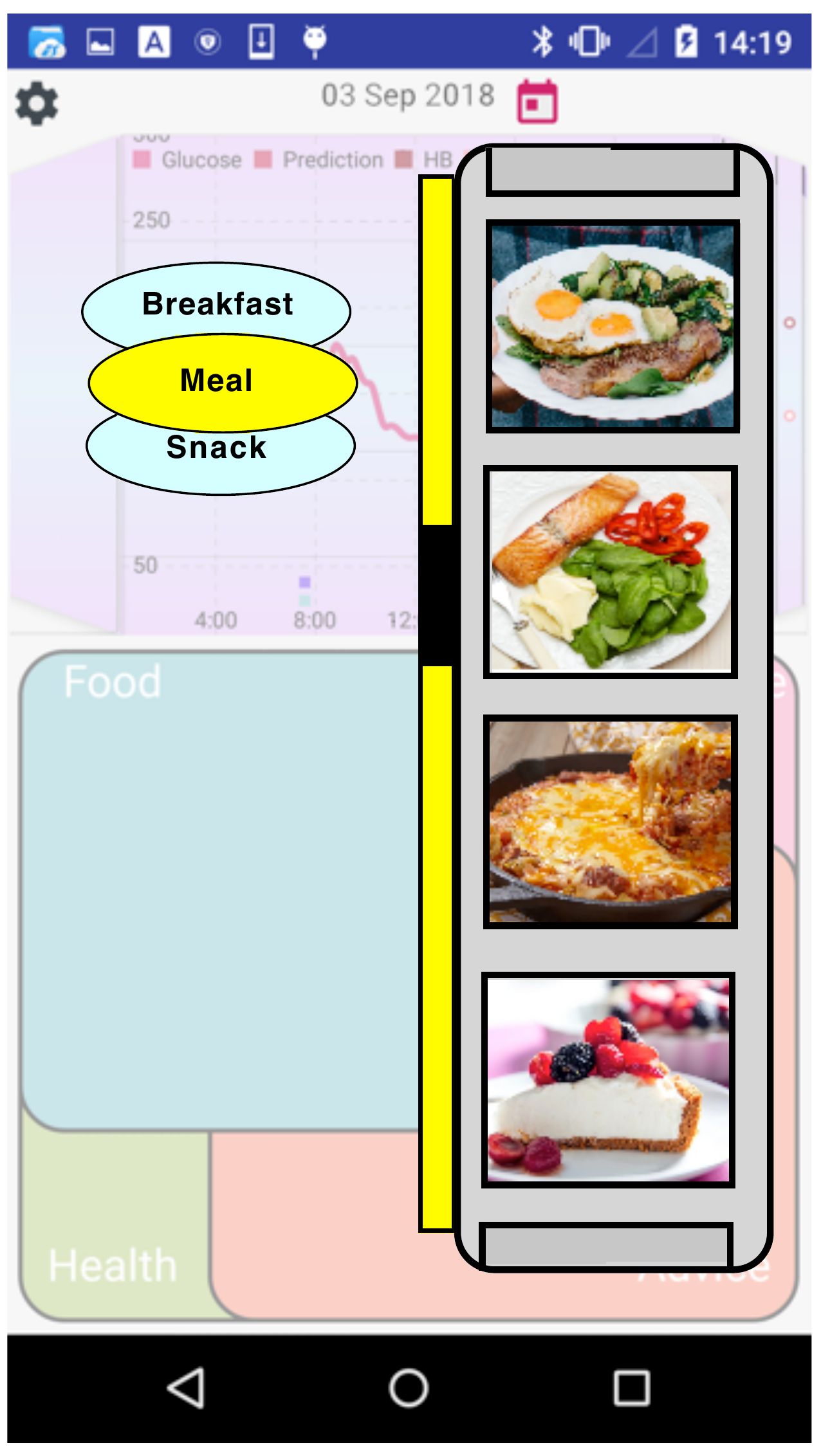}
  \captionof{figure}{A scrollable menu of favourite meals. \\ \mbox{}\\}
  \label{fig:test10}
\end{minipage}%
\ \ \
\begin{minipage}{.49\textwidth}
  \centering
  \includegraphics[height=4in]{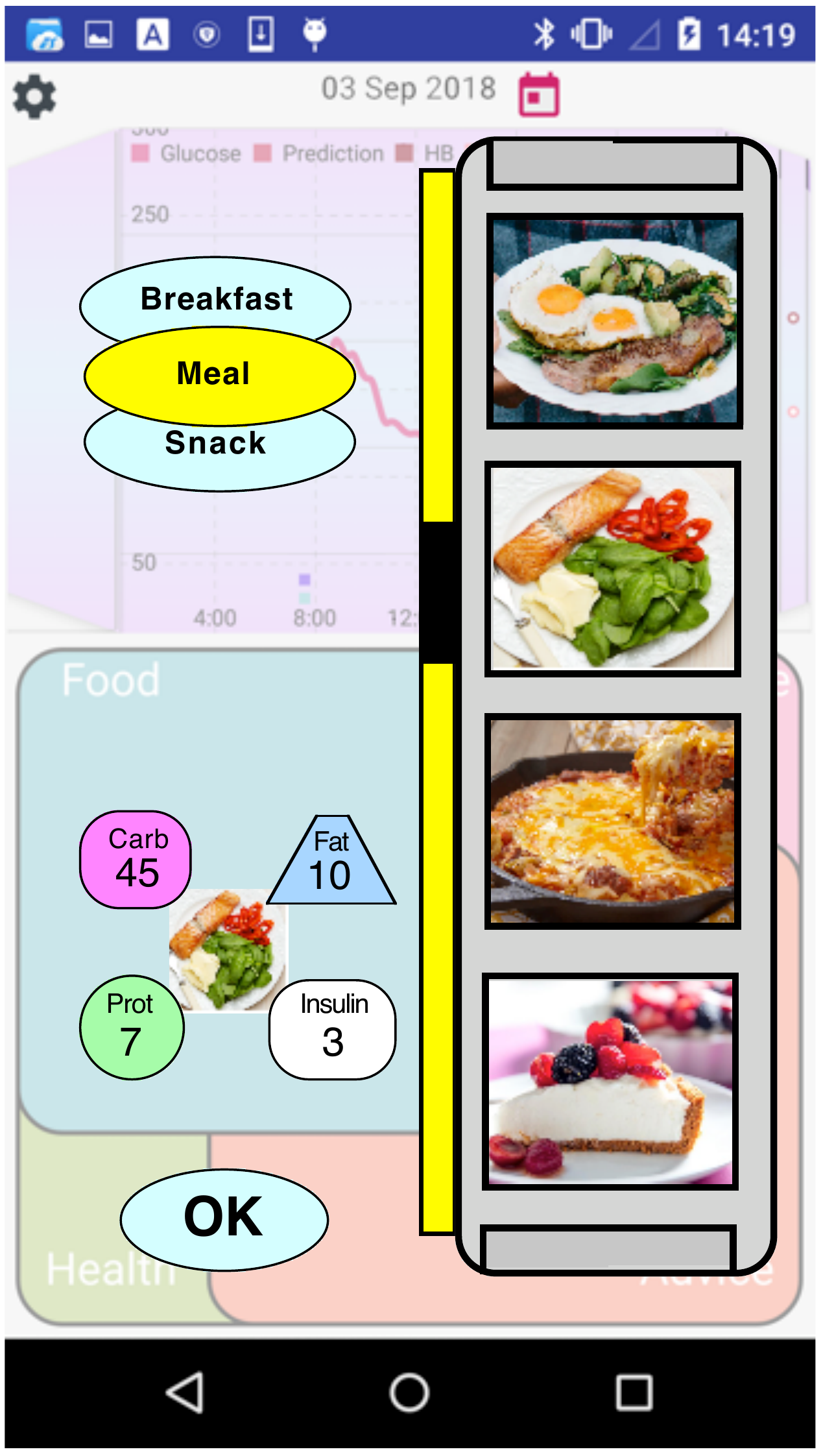}
  \centering
  \captionof{figure}{Touch selection of a meal causes its macronutrient values and recommended insulin dose to be presented for confirmation.}
  \label{fig:test11}
\end{minipage}
\end{figure}



\begin{figure}
\centering
\begin{minipage}{.49\textwidth}
  \centering
  \includegraphics[height=4in]{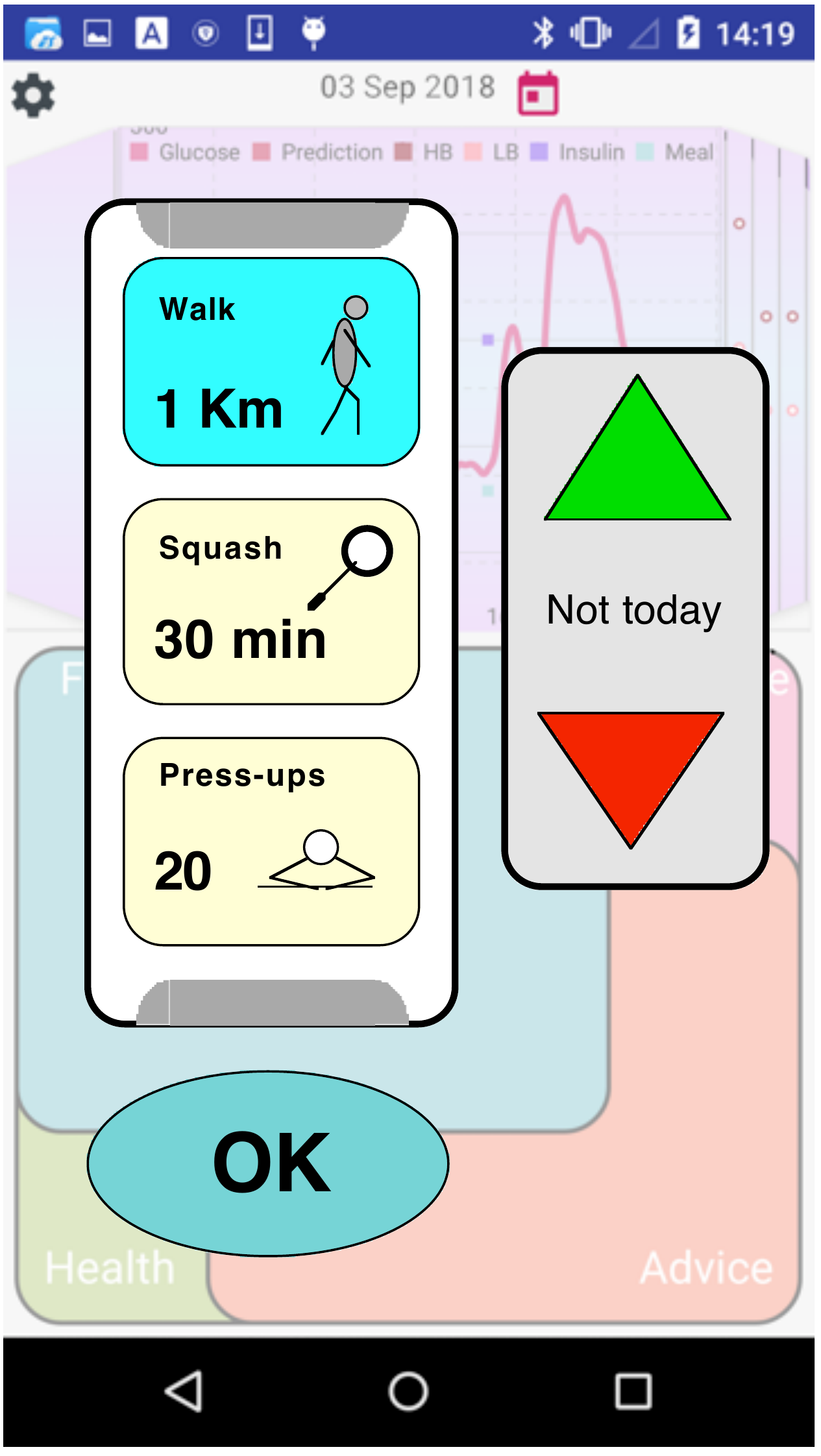}
  \captionof{figure}{Exercise menu allowing selection of type and intensity.\\ \mbox{} \\ \mbox{}\\\mbox{}\\}
  \label{fig:test12}
\end{minipage}%
\ \ \
\begin{minipage}{.49\textwidth}
  \centering
  \includegraphics[height=4in]{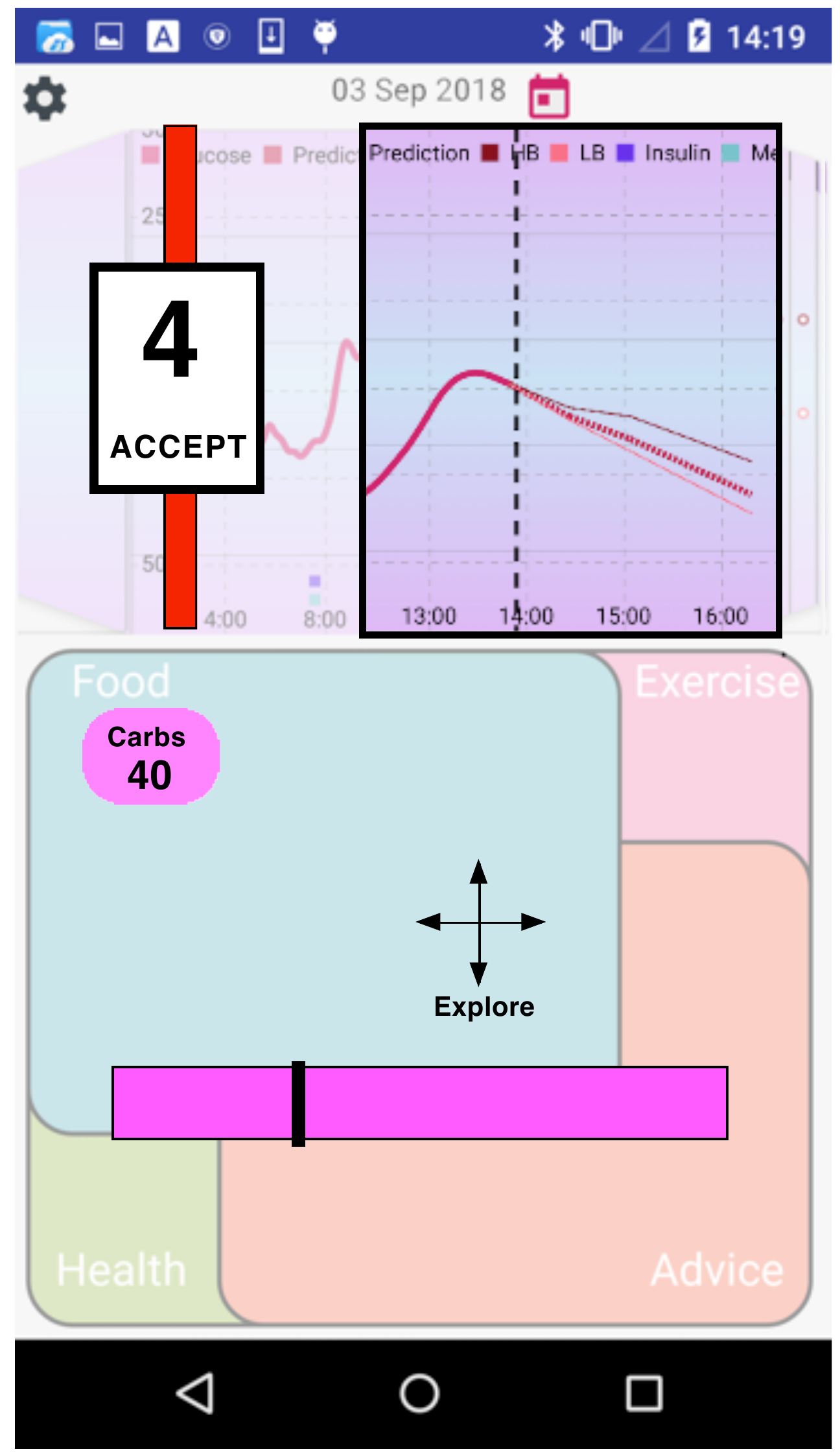}
  \centering
  \captionof{figure}{Interface appearance following a touch on the Explore icon in the Food region. A moving touch on the carb slider, currently showing a value of 40, leads to the display of corresponding predicted and confidence  limits of blood glucose as well as the recommended insulin dose. A second touch on the Explore icon causes the last carb value to be recorded as selected, and a return to the state. of which Figure 3 is a part example.
  }
  \label{fig:test13}
\end{minipage}
\end{figure}



In response to touch on the Explore icon in the Food region, the interaction mechanism shown in Figure 13 appears and supports dynamic exploration of the effect of carb value on predicted blood glucose and recommended insulin dose. The bolus value of the latter (here 4) appears in the box on the left. At top right is a plot of recorded blood glucose up to the present time as well as its prediction and related confidence limits for the next two hours for the default value (here 40) of carb shown in the Food region. The horizontal slider allows scrolling to simulate variation in carb value below and above the default value, causing the predicted blood glucose value and the recommended insulin dose to change appropriately. In this way, for the planning of a meal, the user can make an informed choice of carb value which then appears in the Food region. Touch on the Explore icon returns the app to its launch state (Figure 8). Using a similar sequence, a finger-induced vertical movement of the insulin dosage box can be brushed, for a given carb value, into the predicted blood glucose plot.

Dynamic exploration is capable of generalization to any parameter that influences blood glucose or any other dependent variable. It could, for example, support exploration of the relationship between caffeine (in the Food region) and potential warnings about hypoglycaemia (in the Advice region) an example of the benefit of having fractions of the four regions (Food, Exercise, Advice, Health) always visible.

\section{Discussion}

The design exercise described above is reported because (1) the opportunities it exploits may well be relevant to a wide range of interactive applications, not only to the management of chronic medical conditions, and (2) it has resulted in an anticipated  powerful and comprehensive support for people with diabetes.

The design emphasises the importance of three key features (1) the beneficial role that appropriately designed context presentation can play; (2) the fact that an understanding of  a user's focus of attention during decision making can lead to the release of almost the entire display screen for the presentation of  menu options; and (3) the potential, for a decision support interface, made possible by machine learning \cite{hayeri-predicting2018, Li-ADLPlatform2018,Perez-DecisionSup2018} supported by wearable physiological sensors, allowing ``What if?'' questions to be explored by a user. Other anticipated advantages of the interface include the ability of the bifocal diary to permit both quantitative and qualitative \cite{Spence-DataBaseNav1982,Mackinlay-ConeTrees1991,Card-Comm2012} awareness of items of temporal interest spread over many days, as well as an anticipated avoidance of typical navigational problems \cite{spence-framework1999,Jul-navigation1997,field-context1990}, partly due to the fact that no major information item ever completely disappears from view.

Under the generic title of Fluid Interaction, Elmqvist et al \cite{Elmqvist-FluidInt2011} have drawn attention to interaction guidelines which, if followed, can substantially enhance the usability of an interface. Their detailed practical recommendations leading to effective interaction experiences have been followed. An important one addresses the common problem of instantaneous response to interaction. Such an effect can be disadvantageous for two reasons. One arises from that feature of the human visual system known as Change Blindness \cite{Rensink-ToSeeOrNot1997}. Another \cite{Mackinlay-ConeTrees1991} is due to the well-established fact that, if a visible change occurs instantaneously, considerable cognitive effort may be needed to update the user's mental model of a system. For these two reasons the implementation of the ARISES app interface ensures that the effect of any interaction/touch is animated smoothly over a period of around 150 to 400 milliseconds.

\section{Conclusion}
A subjective conclusion derived from regular meetings alongside people with diabetes is that the anticipated benefits of the novel aspects of the ARISES app may well be realised. Objective conclusions will emerge from an extensive formal clinical study scheduled to start soon on the completion of the design's implementation.

\section{Future Work}

Following the implementation and first testing of the ARISES app it will be evaluated. Evaluations of existing hand-held devices (e.g. ABC4D \cite{Herrero-AdvInsBolus2015}, and GlucOracle \cite{desai-personal2019}) have been reported  and illustrate the kind of methodologies that can be employed. In a complementary approach we plan to investigate numerical aspects of the ARISES app such as the excess of interactions over the minimum required for specific tasks, the time taken to perform tasks, and task performance on first acquaintance without explanation \cite{de-m-rsvp2004,Budiu-Interaction2013,lam-framework2008}. The outcomes of such experimental observations, both subjective and objective, will inform redesign of the app.

\section{ACKNOWLEDGEMENTS}
The support received from Monica Reddy, Taiyu Zhu, Ryan Armiger and Jugnee Navada is gratefully acknowledged. The research is supported by EPSRC Grant EP/P00993X/1.

\section{APPENDICES}

\textbf{A1 \ Clinical \ background }\\
Type 1 diabetes is a chronic condition requiring regular blood glucose monitoring and adjustment of insulin dosage according to activities of daily living such as eating and exercise.  Continuous glucose monitoring sensors and insulin bolus calculators are existing technologies \cite{garg-improved2008,lepore-bolus2012,ziegler-use2013,pesl-advanced2015,vallejo-calculating2017} proven to facilitate self-management by improving glucose control and the reduction of hypoglycaemia associated with excess insulin dosage. Continuous glucose monitoring devices use a subcutaneously deployed sensor and Bluetooth technology to feed back real-time interstitial fluid glucose levels to the user's smartphone. Insulin bolus calculators are increasingly being integrated into glucose metering software and use patient entered carbohydrate content to calculate an appropriate insulin dose corresponding to food intake. The majority of diabetes health apps provide treatment decision support by presenting automatic and manually entered data inputs from these existing diabetes technologies in a meaningful fashion. Arsand et al \cite{Arsand-MobHeaApp2012} used various end-user-based assessments to evaluate the functionality of ten diabetes mobile health app features and outlined key components for future app development. Important features included: use of automatic data transfer; motivational and visual interface design; greater health benefit-to-effort ratio; dynamic usage; and applying context to app output. Herrero et al \cite{Herrero-AdvInsBolus2015} analysed how various human and environmental factors such as exercise, stress and alcohol consumption, influence blood glucose levels and how these parameters can be incorporated in intelligent decision support systems.

A review of commercially available diabetes health apps \cite{Wu-MobileApp2017} established that having a structured display was a feature that significantly improves blood glucose control. This association is likely a result of positive health behavior changes in response to well-presented health outcomes (e.g. blood glucose data). A randomized cross-over study investigating glucose prediction as part of a diabetes decision support system revealed further decision modification in $20\%$ of cases during the intervention arm \cite{Perez-DecisionSup2018}. The presentation of structured glucose prediction data generated from an artificial intelligence (AI) with self-learning capabilities together with the ability to take account of real-time physical activity, provides an opportunity to engage the user and further improve clinical outcomes. The presence of educational and lifestyle modification features are also low risk additions that increase self-awareness and improve glucose control \cite{Wu-MobileApp2017}.

The T1DM population comprises people from different cultural and generational backgrounds with varied ideas about, and experience with, the use of mobile health applications. Interface design for most diabetes decision support apps struggle to efficiently permit layered multi-source data inputs and present various outputs (e.g. blood glucose and insulin dose recommendation) without compromising ease of use, context and the number of device interactions. The ARISES app described in this paper aims to overcome these hurdles by adopting evidence derived from the current literature \cite{christie-consultation2015, peiris-going2018,Ben-Priciples2015} and by including people with T1DM with varied exposure to technology within the design activity.

\textbf{A2 \ The \ human \ visual\  system\ and\ peripheral\ sensory\ nervous\ system}\\
The effectiveness of user interaction with a hand-held device is crucially dependent on human visual acuity and peripheral tactile sensation, both of which must be addressed.

Diabetic retinopathy and peripheral sensory neuropathy are two common diabetes related complications that compromise visual acuity and peripheral tactile sensation respectively. Impaired colour vision is another visual complication which can also occur in individuals with and without established retinopathy \cite{Gella-Impairment2015, Tan-FactorsAss2017}. Most common cases of impaired colour vision affect the red-green axis but cases of blue-yellow impairment (tritanopia) have been identified in diabetes populations \cite{Wu-MobileApp2017}).  To overcome neuropathy our interface will utilize the haptic vibration feature within smart-phone hardware. The ability to scale text; the use of customizable pictures and icons; and the avoidance of high contrast colours (red, green, blue and yellow) will all support accessibility to users having impaired colour and visual acuity.

\textbf{A3 \ User\  Preferences}\\
At a series of regular focus meetings T1DM participants strongly recommended that, to avoid therapeutic errors, current blood glucose levels, insulin dose recommendations and insulin-on-board (insulin still active in the body) are clearly and unambiguously presented within the interface.
Most continuous blood glucose sensor applications will alert users when the level falls outside a target range, but the majority of people with diabetes can anticipate, and are sensitive to, physiological symptoms that occur when blood glucose levels deviate to extremes. The ability to modify alert thresholds for blood glucose, and provide advisory support were vital functions suggested to avoid user fatigue.

The identification of four principal input data types (Food, Exercise, Health and Advice) was agreed within the focus meetings, with an emphasis on ensuring that important data associated with each type could remain visible as interaction proceeds.

Focus groups also confirmed and extended our understanding of habitual behaviour to meal choices and physical activity. Variety in dietary choices are often limited within the constraints of stable dietary intake behaviour determined by multiple factors including eating habits and core food environment. This stable behaviour results in choices within a set food category with minimal impact on core dietary behaviour \cite{mela-Food1999}. The inclusion of features that allow the patient to save and choose previously entered meal profiles via the display of meal images can exploit this habitual dietary behaviour and make the process of meal selection easy and efficient. The same concept is transferable to the input of exercise choices.


The ability to review the glycaemic outcomes of historic events was considered to be an important feature in the support of future treatment decisions. For example, the incorporation of a customizable event and data filter in the Heath region, and using a diary to visually review the circumstances surrounding blood glucose outcomes, delivers an ingeneous level of precision and context in presenting historical data. Such a feature not only indicates when specific events and outcomes occur, but helps to explain why and what actors could have contributed. A general statistics pop-up accessible in the Health domain will provide users with a welcome snapshot of daily, weekly or monthly data (e.g., daily insulin requirements, percentage time in glycaemic range, frequency of hypoglycaemia and exercise data).  In addition, intercurrent illness and stress are well known factors which cause an increase in blood glucose \cite{Lloyd-Stress2005} . Therefore, the addition of a stress and illness switch was suggested to alert the AI to make adaptations to compensate for any fluctuations as a result of these. A clearly visible icon on the diary will serve as a reminder to the user to switch this feature off following recovery.



%

\bibliographystyle{ACM-Reference-Format}

\end{document}